\documentclass[prl,superscriptaddress,twocolumn,showpacs]{revtex4}
\usepackage{graphicx} 
\usepackage{amsmath}
\usepackage{amsfonts}

\newcommand{\ket}[1]{\left|#1\right\rangle}
\newcommand{\bra}[1]{\left\langle#1\right|}
\newcommand{\braket}[2]{\left\langle#1|#2\right\rangle}

\begin{document}

\title{Measuring ultrasmall time delays of light by joint weak measurements}
\author{Gr\'egory Str\"ubi}
\affiliation{Department of Physics, University of Basel, 
Klingelbergstrasse 82, CH-4056 Basel, Switzerland}
\author{C. Bruder}
\affiliation{Department of Physics, University of Basel, 
Klingelbergstrasse 82, CH-4056 Basel, Switzerland}

\begin{abstract} We propose to use weak measurements away from the
  weak-value amplification regime to carry out precision measurements
  of time delays of light. Our scheme is robust to several sources of
  noise that are shown to only limit the relative precision of the
  measurement. Thus, they do not set a limit on the smallest measurable phase
  shift contrary to standard interferometry and weak-value based
  measurement techniques. Our idea is not restricted to phase-shift
  measurements and could be used to measure other small effects using a
  similar protocol.
\end{abstract}

\pacs{03.65.Ta, 42.50.-p, 07.60.Ly}


\maketitle

Twenty years after the proposal \cite{Aharonov1988} of Aharonov,
Albert, and Vaidman, several experiments have demonstrated the
possibility to measure tiny physical effects using the so-called
weak-value amplification scheme~\cite{Hosten2008,Dixon2009,
  Gorodetski2012}. These ideas have paved the way for new approaches
to precision measurements in general, and have triggered a great deal
of further theoretical and experimental developments.

Recently, Hosten and Kwiat~\cite{Hosten2008} were able to
experimentally confirm the spin Hall effect of light by measuring a
polarization-dependent displacement of a laser beam to a precision of
1 \!\AA\quad using weak-value amplification. Gorodetski \textit{et
  al.}~\cite{Gorodetski2012} investigated the plasmonic spin Hall
effect.  Dixon \textit{et al.}~\cite{Dixon2009} determined the angle
of a mirror to a precision of the order of 500 frad by measuring
deflection of light off the mirror. Several experiments were proposed
in order to enhance the precision of the measurement of longitudinal
phase shifts of light~\cite{Brunner2010}, or amplify the single-photon
nonlinearity to a measurable effect~\cite{Feizpour2011}. An
application to charge sensing in a solid-state context has also been
put forward~\cite{Zilberberg2011}. The advantages of weak-value
amplification schemes for suppressing technical noise were
investigated in~\cite{Feizpour2011, Starling2009, Nishizawa2012},
and~\cite{Kedem2012} showed how technical noise could even improve the
precision in this scheme. Ways to optimize the initial meter
wavefunction were studied in~\cite{Susa2012}.

There was a number of attempts to go beyond the
weak-value formalism. This includes, for instance, higher-order
expansions for nearly orthogonal pre- and post-selected states~\cite{Geszti2010, Wu2011, Nakamura2012, Pang2012, Kofman2012}, or
the use of full counting statistics~\cite{DiLorenzo2012}, and
orbital-angular-momentum pointer states~\cite{Puentes2012}. The effect
of decoherence was investigated in Ref.~\cite{Knee2013}. The
limits of amplification for arbitrary coupling strength were addressed
in~\cite{Koike2011, Zhu2011}. Connections of the weak-value formalism
with the theoretical tools of precision metrology were made
in~\cite{Hofmann2011, Hofmann2012}. Weak values were also related to 
quasiprobability distributions of incompatible observables~\cite{Bednorz2012}.

In this Letter, we would like to propose a scheme to measure very
small time delays, or longitudinal phase shifts. The idea is to use
weak measurements away from the weak-value amplification regime and to
exploit the full information contained in the correlations induced by
the time delay between frequency and polarization of photons going
through the interferometer. This procedure is not limited to
time-delay measurements and could be used for other precision
measurements. For example, it is readily applicable to measurements of
ultrasmall beam deflections by slightly modifying the protocol of
Dixon \textit{et al.} \cite{Dixon2009}. The idea of carrying out a
full joint measurement of two weakly entangled degrees of freedom
could be relevant in many domains such as charge sensing in solid
state physics \cite{Zilberberg2011}, precision metrology, and
gravitational wave detection.

Following Ref.~\onlinecite{Brunner2010} we are interested in measuring
with high precision a small interaction parameter $\tau$ which couples
two physical systems.  We will call them \emph{system} and
\emph{meter} as is customary in weak-value protocols, and the effect
of the coupling is described by an unitary operator
\begin{equation} 
\hat U = e^{i \tau \hat A \hat x}\,,
\label{eq:interaction}
\end{equation} 
where $\hat A$ is an operator acting on the system and $\hat x$ acts
on the meter.
There are many different situations where the measurement of a small
coupling constant $\tau$ is potentially relevant. For example the
interaction can be an interesting but very small physical effect, like
the spin Hall effect of light \cite{Hosten2008}; or $\tau$ can be a
quantity of direct interest, describing the angle of a mirror, see
\cite{Dixon2009}, or describing the amplitude of a gravitational wave
in an interferometric setup. In this Letter we will consider a setup
where a time delay in an optical interferometer plays the role of the
interaction parameter $\tau$, as in~\cite{Brunner2010}.

The interaction~(\ref{eq:interaction}) with small $\tau$ is formally
similar to a weak measurement of the observable $\hat A$ of the
system. Therefore by adding a \emph{pre-} and \emph{post-selection} of
the system in states $\ket i$ and $\ket f$ respectively,
see~\cite{Aharonov1988}, we obtain the weak value of the observable
$\hat A$, $A_w= \bra f \hat A \ket i/\braket f i$ in the meter
averages 
$\langle\hat p\rangle = \tau \text{Re}A_w$, and $\langle \hat x
\rangle = -2 \tau \bra\psi\hat x^2\ket\psi \text{Im}A_w$, where
$\ket\psi$ is the initial state of the meter. Those results are only valid up to linear order in $\tau$, which is enough for our purposes, but the meter averages can also be expressed exactly using a generalized form of weak values~\cite{Dressel2012}. Weak-value amplification
is based on the fact that it is possible to obtain arbitrarily large
weak values by choosing almost orthogonal initial and final states,
$\braket f i\rightarrow 0$, thereby \emph{amplifying} the small
interaction parameter $\tau$. The price to pay for obtaining large
weak values is the small probability $P_\text{ps} = \left|\braket f i
\right|^2$ of a successful post-selection. Weak-value amplification
does not increase the statistical information because the
amplification of the signal is exactly counterbalanced by the small
success rate of the post-selection. However the amplification allows
to overcome technical limitations, {\it e.g.} detector resolution or
readout noise. On the other hand, stringent post-selections lead to an
increased sensitivity against certain types of noise. For instance
background noise, {\it e.g.} stray photons hitting the detector in
optics experiments, is a limiting factor of how small the
post-selection can be since the rate of useful, post-selected, photons
must be significantly larger than the background. Another critical
type of noise is due to errors in the post-selection itself.

Conceptually, as is the case for all quantum mechanical
measurements~\cite{Dressel2010}, the crucial information in imaginary
weak-value amplification is not the average shift of the meter, but
the correlations induced by the interaction between the system and the
meter. These are generally probed by joint measurements on the meter
and system. Weak-value amplification constitutes a partial joint weak
measurement, since measurement outcomes are retained only if the
system is found in a certain state. Remarkably, it turns out that all
the useful correlations for the estimation of $\tau$ are contained in
the small post-selection regime. However, additional information on
other aspects of the experiment is included in the other outcomes. In
the presence of noise, this additional information turns out to be
useful to precisely evaluate $\tau$. In the following, we will present
an example where carrying out a full joint weak measurement
outperforms weak-value amplification.

\begin{figure}
\centering
  \includegraphics[width=0.9\linewidth]{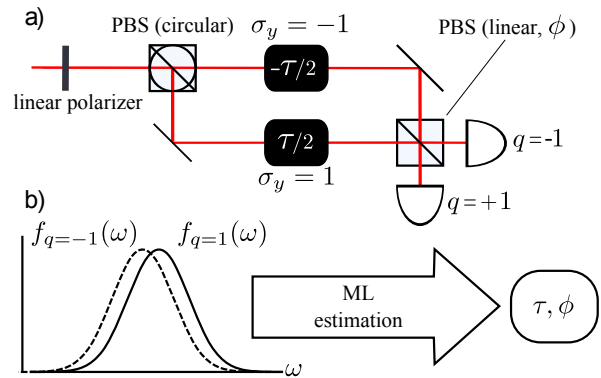}
  \caption{(a) Mach-Zehnder interferometer for laser light with a
    time-delay asymmetry $\tau$. The incoming beam is linearly
    polarized and then evenly split by a polarizing beam splitter
    (PBS) into the two arms with opposite circular polarizations,
    $\sigma_y=\pm1$. The beams are recombined at a second PBS with
    linearly polarized outputs. The direction of the linear
    polarization is described by the azimuthal angle $\phi$ on the
    Poincar\'e sphere. Two spectrometers labeled by $q=\pm1$ measure
    the spectrum of the outgoing light. (b) Graphical representation
    of the estimation process. The values of $\tau$ and $\phi$ are
    extracted from the spectra measured by the two detectors by a
    maximum-likelihood (ML) estimation technique as described in the
    text.}
  \label{fig1}
\end{figure}

\textit{Time-delay measurements}.-- 
We consider a Mach-Zehnder interferometer for laser light, see
Fig.~\ref{fig1}a.  Encoding the which-path information in a two-level
system allows us to write the effect of the time delay as
\begin{equation} 
U = e^{i\tau \hat \sigma_y \hat H/2}\,,
\end{equation} 
where $\hat \sigma_y$ acts on the which-path space, and $\hat H$ is the
Hamiltonian of the laser light. The time delay now takes the role of
the interaction parameter in Eq.~(\ref{eq:interaction}), 
and the photon frequencies (which are contained in $\hat H$) the role
of the pointer variable $x$. The incoming
light is evenly split into the two arms with opposite circular polarizations by a polarizing beam
splitter. The two arms then recombine at another polarizing beam
splitter with linearly polarized outputs. The direction of the linear polarization is described by the azimuthal angle $\phi$ on the Poincar\'e sphere. The two output ports $q=\pm1$ are monitored by spectrometers. In the weak-value language, the pre-selected polarization state is $\ket i = (\ket - + \ket +)/\sqrt{2}$ and the two post-selected states at detector $q$ are $\ket {f_q} = (\ket - + q e^{i\phi}\ket +)/\sqrt2$. We
are interested in ultrasmall time delays with
$\omega_\text{max}\tau\ll1$, where $\omega_\text{max}$ is the maximal
frequency contained in the laser light.

A laser pulse with normalized spectrum $p_0(\omega)$, which can be
interpreted as a probability density, is sent through the
interferometer. The probability density of outcomes is then given by
\begin{equation} 
p_q(\omega;\tau,\phi) =
\frac12p_0(\omega)
[1+q\exp(-\epsilon^2/2)\cos(\phi-\omega\tau)]\,,
\label{eq:probability_density}
\end{equation} 
where we have introduced Gaussian fluctuations of amplitude $\epsilon$
of the alignment parameter $\phi$, \textit{i.e.}, convoluted the bare
probability density with a fluctuation kernel
\begin{equation}
\xi(\phi,\phi') = \frac{1}{\sqrt{2\pi}\epsilon}
\exp\left(-\frac{(\phi-\phi')^2}{2\epsilon^2}\right)  
\label{fluctuation_kernel}
\end{equation}
that describes fluctuations around the average alignment $\phi$.
Note that only the fluctuations with a correlation time smaller than
the measurement duration are included in $\epsilon$. Fluctuations with
a longer correlation time will modify the effective value of
$\phi$. From now on we make the realistic assumption $\epsilon \ll 1$. It can be shown that fluctuations of the alignment of the circular polarizing beam splitter and of the linear polarizer can also be encompassed in $\epsilon$.

Our goal is to find an estimate of the values of the parameters $\tau,
\phi$ from outcomes of an experiment following the probability
distribution~(\ref{eq:probability_density}). Although $\phi$ is in principle controlled experimentally, its estimation permits to remove possible systematic errors. The quantity produced by
the experiment is a set of observed frequencies $f_q(\omega)$, which,
in principle, converge to $p_q(\omega;\tau,\phi)$ as the number $N$ of
measured photons is increased.

We use the maximum likelihood procedure~\cite{Helstrom1976} to 
provide unbiased estimates
of the values of the parameters, see Fig.~\ref{fig1}b. This is achieved by maximizing the
log-likelihood function $l(\tau,\phi)$ defined as
\begin{equation} l(\tau,\phi) = \sum_{q=\pm1}\int d\omega\; f_q(\omega)\log
p_q(\omega;\tau,\phi)\,.
\end{equation} 
Maximizing the log-likelihood yields two equations which have to be
solved numerically in the general case. However, for the special case
of almost equal intensities at the two detectors, $|\phi-\omega\tau -
\frac\pi2| \ll 1$, analytical expressions can be derived
\begin{align}
\phi &= \frac\pi2 - \exp(\epsilon^2/2)\sum_q qP_q\,,\label{eq:phi_estimate}\\
\tau &= \frac1{4\Delta\omega}\exp(\epsilon^2/2)\left(\frac 1{\Delta\omega}\sum_q
qP_q \langle \omega\rangle_q-\sum_q qP_q\right) \,,
\label{eq:tau_estimate}
\end{align} 
where $P_q$ is the integrated fraction of outcomes in detector $q$,
and $\langle\cdot\rangle_q$ denotes the average value in detector
$q$. The frequency spread $\Delta\omega$ of the initial distribution
is given by $(\Delta\omega)^2 = \sum_q P_q\langle \omega^2\rangle_q
-\left(\sum_q P_q \langle \omega\rangle_q\right)^2$. The
estimates~(\ref{eq:phi_estimate},\ref{eq:tau_estimate}) depend on measurement results and on
one unknown amplitude $\epsilon$ that characterizes alignment fluctuations.
Remarkably, $\exp(\epsilon^2/2)$ only appears as an overall multiplicative
factor. Thus, the ultimate error $\Delta\tau_\text{ult}$ on the estimation of $\tau$ by assuming $\epsilon=0$
(since its value is unknown, yet realistically $\epsilon\ll1$) scales with
$\tau$, \textit{i.e.} only a relative error occurs that does not
limit the smallest $\tau$ that can be measured.
Equation~(\ref{eq:tau_estimate}) is one of the main results of our
Letter.

Aside from the errors due to technical noise, statistical uncertainties also contribute to the estimation error. For a finite number $N$ of detected photons, the statistical errors are provided by the Cram\'er-Rao
bound~\cite{Helstrom1976}: 
$ \Delta\tau_\text{stat} \geq \sqrt{(\mathcal I^{-1})_{\tau\tau}/N}$, 
and $ \Delta\phi_\text{stat} \geq \sqrt{(\mathcal I^{-1})_{\phi\phi}/N}$, 
where $\mathcal I$ is the Fisher information matrix given by
\begin{equation} 
\mathcal I_{y z} = \sum_{q=\pm1}\int dx\;
p_q(x;\tau,\phi)\left(\partial_y\log p_q\right)\left(\partial_z\log
p_q\right)\,.
\end{equation}
As a side remark, note that the logarithmic derivative $\partial_\tau \log p_q$ is connected to the weak value of $\sigma_y$ in detector $q$, see Ref.~\onlinecite{Hofmann2012} for a discussion.
Asymptotically, {\it i.e.}, for large $N$, the Cram\'er-Rao bounds are
saturated by the maximum-likelihood procedure. In the case of almost
equal intensities at the two detectors we obtain
\begin{equation} 
\Delta\tau_\text{stat} \geq\frac{\exp(\epsilon^2/2)}{4\Delta\omega\sqrt{N}}\,,
\label{eq:tau_statistical_noise}
\end{equation} 
which shows that the fluctuations of $\phi$ do not increase
significantly statistical noise. The number of detected photons $N$
required to obtain a good estimate of $\tau$ is of the order
$10/(\Delta\omega\tau)^2$. Estimating a time delay of the order of $1$
attosecond with ultrashort laser pulses with
$\Delta\omega\approx10^{15}\text{Hz}$ would require detecting $10^7$
photons. A typical pulse contains typically $10^{13}$ photons which
would be enough to measure time delays of the order of $1$ zeptosecond
corresponding to a displacement of 100fm.

\textit{Split detectors}.-- 
We would now like to apply this result to an actual detector with a
finite resolution. This will in particular shed light on the roles
played by the resolution and readout noise of the detector. In
principle, the results of the previous section could require measuring
the full distribution of $\omega$ without any readout noise. To show
that this is not the case we consider ``split'' detectors which can
only discriminate two spectral regions $\omega > \omega_0$, leading to
a measurement result $r=+1$, and $\omega < \omega_0$, leading to
$r=-1$. We also add readout noise from the outset. We will see that
our conclusion from Eq.~(\ref{eq:tau_estimate}) (\textit{viz.} that
there is no absolute lower limit on the smallest value of $\tau$ that
can be measured) survives even in this extreme case.

To make up for losing the detector resolution some \textit{a priori}
knowledge of the initial frequency distribution $p_0(\omega)$ is
required. For analytical calculations we will assume an initial
Gaussian distribution, but the results will depend only quantitatively
on the shape. In practice the distribution should be measured and the
calculations done numerically. We thus assume $p_0(\omega) \approx
e^{-(\omega-\omega_0)^2/2(\Delta\omega)^2} / \sqrt{2\pi}\Delta\omega$
where the tails should, in principle, be truncated since
$0\leq\omega\leq\omega_{\text{max}}$. The probabilities of the two
possible outcomes $r=\pm 1$ at the two detectors $q =\pm 1$ at second
order in the small quantity $\Delta\omega \tau$ read
\begin{align} 
p_{rq} = \frac14[1+q\exp(-\epsilon^2/2)
(1-\frac12(\Delta\omega\tau)^2)\cos\phi] \nonumber\\
+rq \exp(-\epsilon^2/2)
\frac{\Delta\omega\tau}{2\sqrt{2\pi(1+(\Omega/\Delta\omega)^2)}}\sin\phi\,,
\end{align} 
where we allowed for a frequency detection uncertainty
of order $\Omega$ modeled as Gaussian white noise. 

Denoting the measured probabilities by $f_{rq}$, the maximum likelihood
estimation can be analytically carried out in the regime
$|\sin\phi|\gg\Delta\omega\tau$, and yields
$
\cos\phi = \exp(\epsilon^2/2)(P_+-P_-)
$
in terms of the integrated fraction of outcomes 
$P_q = \sum_r f_{rq}$, and
\begin{equation} 
\tau = \frac{\sqrt{2\pi(1+(\Omega/\Delta\omega)^2)}}
{8\Delta\omega\sqrt{\exp(-\epsilon^2)-(P_+-P_-)^2}}
\sum_{r,q}\frac{rqf_{rq}}{1+q(P_+-P_-)}\,.
\label{eq:tau_estimate_split}
\end{equation} 
The first terms in the expansion of $\tau$ in the
fluctuations $\epsilon$ and detection noise $\Omega$ 
can now be calculated, and we finally obtain
\begin{equation} 
\tau =\tau_0\left[1+\frac12\left(\frac{\epsilon}{\sin\phi}\right)^2+
\frac{1}{2}\left(\frac{\Omega}{\Delta\omega}\right)^2\right]\,,
\label{eq:tau_estimation_split_expansion}
\end{equation} 
where $\tau_0$ is the estimated value of $\tau$ in the absence of
noise, see Eq.~(\ref{eq:tau_estimate_split}) with $\epsilon$ and $\Omega$ set to zero. Again the noise does not set an absolute limit on the precision
of the estimation of $\tau$ but only a relative precision. The frequency spread $\Delta\omega$ reduces the effect of readout noise. Moreover we
observe that in order to minimize the effect of fluctuations of
$\phi$, we have to work in the regime $\sin\phi\approx1$ and not in
the weak-value amplification regime of $\sin\phi\approx0$ where the
effect of fluctuations increased.

The statistical uncertainty is given by the Fisher information matrix,
which is diagonal in this case, through the Cram\'er-Rao bound
\begin{equation} 
\Delta\tau_\text{stat} \geq \frac{\sqrt{2\pi}}{4\Delta\omega\sqrt{N}}\left[1+
\frac12\left(\frac{\epsilon}{\sin\phi}\right)^2+
\frac12\left(\frac{\Omega}{\Delta\omega}\right)^2\right]\,.
\label{eq:tau_estimation_split_error}
\end{equation} 
Hence the considered fluctuations do not jeopardize the estimation of
$\tau$ away from the weak-value amplification regime. We also note
that using split detectors leads to a modest increase of statistical
noise by a factor $\sqrt{2\pi}$ with respect to
Eq.~(\ref{eq:tau_statistical_noise}).
Equations~(\ref{eq:tau_estimation_split_expansion},\ref{eq:tau_estimation_split_error})
demonstrate that even for low-resolution detectors our scheme is
robust against readout noise and alignment errors.

\textit{Comparison to existing schemes}.-- 
Standard interferometry compares two probabilities given by the sine
and cosine of the total phase shift given by the combination
$\phi-\omega\tau$. This leads to two difficulties: first, to estimate
$\tau$ precisely, the laser frequency $\omega$ has to be highly
stabilized. Secondly, the alignment $\phi$ cannot be
separated from the effect of $\tau$, \textit{i.e.}, alignment errors
are a limiting factor to the precision achievable having complete
statistical information. This ultimate precision, which cannot be increased by acquiring more measured data, is given by 
$\Delta\tau_\text{ult} = \epsilon/\omega$. 
In the procedure proposed by Brunner and Simon \cite{Brunner2010} 
that uses the imaginary part of the weak value, the first issue is solved
since a large frequency spread $\Delta\omega$ is advantageous for the
precise evaluation of $\tau$, which is also true in our
scheme. However, the second issue is only partially addressed in
\cite{Brunner2010}: alignments errors are still a limiting factor,
$\Delta\tau_\text{ult} = C\epsilon$, where the proportionality 
constant $C\approx 10^{-18}$~s is
three orders of magnitude smaller than for standard interferometry.

\begin{figure}
\centering
  \includegraphics[width=0.9\linewidth]{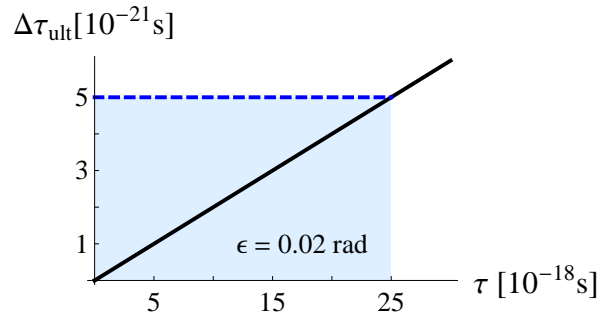}
  \caption{Ultimate precision limit $\Delta\tau_\text{ult}$ on the
    estimation of the physical time delay $\tau$ as a function of
    $\tau$, in the presence of fluctuations of the alignment $\phi$,
    see Eq.~(\ref{fluctuation_kernel}). We assume $\epsilon = 0.02$
    rad for the plot.  The dashed line shows the constant
    (zeroth-order) contribution of fluctuations to
    $\Delta\tau_\text{ult} = C\epsilon$ for weak-value amplification,
    here we assume $C\approx0.25\times10^{-18}$~s like in
    \cite{Brunner2010}. The solid line is the result of our scheme,
    $\Delta\tau_\text{ult} = \epsilon^2\tau/2$. For $\tau <
    2C/\epsilon$ (shaded area), our scheme outperforms weak-value
    amplification .}
  \label{fig2}
\end{figure}

In contrast to that, a major advantage of our scheme is to remove
systematic errors as well as fluctuations of the alignment as a
limiting factor to the ultimate precision of the time-delay
measurement. Alignment fluctuations lead to a relative error
$\Delta\tau_\text{ult} = \epsilon^2\tau/2$ on the estimation of
$\tau$, see the discussion after Eq.~(\ref{eq:tau_estimate}) and the
illustration in Fig.~\ref{fig2}. This is made possible by working away
from the weak-value amplification regime and using all the information
contained in the correlations between frequency and polarization of
the photons to perform a simultaneous estimation of $\phi$ and $\tau$.

Finally, we would like to mention that if the detector can only
measure events at a small rate due to detector saturation, small
post-selection probabilities as realized in weak-value amplification
schemes permit to effectively increase the rate of
measurements~\cite{Starling2009}.  This allows to obtain better
statistics, however, if the measurement is limited not by statistics
but for other reasons, our method appears preferable.

In conclusion, we have proposed a technique that could be useful to
determine very small time delays, or longitudinal phase shifts, due to
its robustness with respect to various noise sources. The time delay
induces correlations between frequency and polarization of photons
going through the interferometer, and our scheme exploits the full
information contained in these correlations.  The idea is not limited
to time-delay measurements and could be used for other precision
measurements.  In a broader setting, the idea of carrying out a full
joint measurement of two weakly entangled degrees of freedom could
find applications in many domains such as charge sensing in solid
state physics, precision metrology, and gravitational wave detection.

\acknowledgments 
We would like acknowledge stimulating discussions with A. Bednorz,
W. Belzig, N. Brunner, and O.~Zilberberg.
This work was financially supported by the Swiss SNF and the NCCR
Quantum Science and Technology.


\begin{thebibliography}{99}

\bibitem{Aharonov1988} Y.~Aharonov, D.~Z.~Albert, and L.~Vaidman,
Phys. Rev. Lett. {\bf 60}, 1351 (1988). 

\bibitem{Hosten2008}
O.~Hosten and P.~Kwiat,  
Science {\bf 319}, 787 (2008).

\bibitem{Gorodetski2012}
Y.~Gorodetski {\it et al}., 
Phys. Rev. Lett. {\bf 109}, 013901 (2012).

\bibitem{Dixon2009}
P.~B.~Dixon, D.~J.~Starling, A.~N.~Jordan, and J.~C.~Howell, 
Phys. Rev. Lett. {\bf 102}, 173601 (2009).

\bibitem{Brunner2010}
N.~Brunner and C.~Simon,  
Phys. Rev. Lett. {\bf 105}, 010405 (2010).

\bibitem{Feizpour2011}
A.~Feizpour, X.~Xing, and A.~M.~Steinberg,
Phys. Rev. Lett. {\bf 107}, 133603 (2011).

\bibitem{Zilberberg2011} 
O. Zilberberg, A. Romito, and Y. Gefen,
Phys. Rev. Lett. {\bf 106}, 080405 (2011). 

\bibitem{Starling2009}
D.~J.~Starling, P.~B.~Dixon, A.~N.~Jordan, and J.~C.~Howell, 
Phys. Rev. A {\bf 80}, 041803 (2009).

\bibitem{Nishizawa2012}
A.~Nishizawa, K.~Nakamura, and M.-K.~Fujimoto,
Phys. Rev. A {\bf 85}, 062108 (2012).

\bibitem{Kedem2012} Y.~Kedem,
Phys. Rev. A {\bf 85}, 060102 (2012).

\bibitem{Susa2012}
Y.~Susa, Y.~Shikano, and A.~Hosoya,
Phys. Rev. A {\bf 85}, 052110 (2012).

\bibitem{Wu2011}
S.~Wu and Y.~Li,
Phys. Rev. A {\bf 83}, 052106 (2011).

\bibitem{Nakamura2012}
K.~Nakamura, A.~Nishizawa, and M.-K.~Fujimoto,
Phys. Rev. A {\bf 85}, 012113 (2012).

\bibitem{Geszti2010}
T.~Geszti,
Phys. Rev. A {\bf 81}, 044102 (2010).

\bibitem{Pang2012}
S.~Pang, S.~Wu, and Z.-B.~Chen,
arXiv:1205.0619(2012).

\bibitem{Kofman2012}
A.~G.~Kofman, S.~Ashhab, and F.~Nori,
Phys. Rep. {\bf 520}, 43 (2012).

\bibitem{DiLorenzo2012}
A.~Di~Lorenzo,
Phys. Rev. A {\bf 85}, 032106 (2012).

\bibitem{Puentes2012}
G.~Puentes, N.~Hermosa, and J.~P.~Torres, Phys. rev. Lett. {\bf 109}, 040401 (2012);
H.~Kobayashi, G.~Puentes, and Y.~Shikano, Phys. Rev. A {\bf 86}, 053805 (2012).

\bibitem{Knee2013}
G.~C.~Knee, G.~A.~D.~Briggs, S.~C.~Benjamin, E.~M.~Gauger,
Phys. Rev. A {\bf 87}, 012115 (2013).

\bibitem{Koike2011}
T.~Koike and S.~Tanaka,
Phys. Rev. A {\bf 84}, 062106 (2011).

\bibitem{Zhu2011}
X.~Zhu {\it et al.},
Phys. Rev. A {\bf 84}, 052111 (2011).

\bibitem{Hofmann2012}
H.~F.~Hofmann, M.~E.~Goggin, M.~P.~Almeida, and M.~Barbieri,
Phys. Rev. A {\bf 86}, 040102(R) (2012).

\bibitem{Hofmann2011}
H.~F.~Hofmann,
Phys. Rev. A {\bf 83}, 022106 (2011).

\bibitem{Bednorz2012}
A.~Bednorz and W.~Belzig,
Phys. Rev. Lett. {\bf 105}, 106803 (2010).

\bibitem{Dressel2012}
J.~Dressel and A.~N.~Jordan,
Phys. Rev. Lett. {\bf 109}, 230402 (2012).

\bibitem{Dressel2010}
J.~Dressel, S.~Agarwal, and A.~N.~Jordan,
Phys. Rev. Lett. {\bf 104}, 240401 (2010).

\bibitem{Helstrom1976}
C.~W. Helstrom,
{\it Quantum detection and estimation theory},
Mathematics in Science and Engineering, vol. 123, (Academic Press, New York, 1976).


\end{thebibliography}
\end{document}